\documentclass[%
 reprint,
superscriptaddress,
 amsmath,amssymb,
 aps,
pra,
]{revtex4-1}

\usepackage{color}
\usepackage{appendix}
\usepackage{braket}
\newcommand{\bg}{\begin{equation}\begin{gathered}}
\newcommand{\eg}{\end{gathered}\end{equation}}
\usepackage{graphicx}
\usepackage{dcolumn}
\usepackage{bm}

\usepackage{overpic}
\begin{document}

\preprint{APS/123-QED}

\title{Discrimination of thermal baths by single qubit probes}

\author{Ilaria Gianani}
 \affiliation{Dipartimento di Scienze, Universit\'a degli Studi Roma Tre, Via della Vasca Navale 84, 00146 Rome, Italy}
 \affiliation{Dipartimento di Fisica, Sapienza Universit\'a di Roma, Piazzale Aldo Moro 5, 00185 Rome, Italy}
 
 \author{Donato Farina}
 \affiliation{NEST, Scuola Normale Superiore, I-56126 Pisa, Italy} 
\affiliation{Istituto Italiano di Tecnologia, Graphene Labs, Via Morego 30, I-16163 Genova, Italy}

 \author{Marco Barbieri}
 \affiliation{Dipartimento di Scienze, Universit\'a degli Studi Roma Tre, Via della Vasca Navale 84, 00146 Rome, Italy}
 \affiliation{Istituto Nazionale di Ottica - CNR, Largo Enrico Fermi 6, 50125 Firenze, Italy}
 
 \author{Valeria Cimini}
 \affiliation{Dipartimento di Scienze, Universit\'a degli Studi Roma Tre, Via della Vasca Navale 84, 00146 Rome, Italy}

 \author{Vasco Cavina}
 \affiliation{Complex Systems and Statistical Mechanics, Physics and Materials Science Research Unit, University of Luxembourg, L-1511 Luxembourg
 }
 
 \author{Vittorio Giovannetti}
 \affiliation{NEST, Scuola Normale Superiore and Istituto Nanoscienze-CNR, I-56126 Pisa, Italy}


\begin{abstract}
Non-equilibrium states of quantum systems in contact with thermal baths help telling environments with different temperatures or different statistics apart. We extend these studies to a more generic problem that consists in discriminating between two baths with disparate constituents at unequal temperatures.
Notably there exist temperature regimes in which the presence of coherence in the initial state preparation is beneficial for the discrimination capability.
We also find that non-equilibrium states are not universally optimal, and detail the conditions in which it becomes convenient to wait for complete thermalisation of the probe.  
These concepts are illustrated in a linear-optical simulation.  
\end{abstract}

\maketitle

\section{Introduction}
%
%

The reduced dynamics of a quantum system interacting with an external environment 
 is typically insensitive to 
many characteristic features of
the latter~\cite{lindblad1976generators,gorini1976completely,petruccione}. 
Yet  some
macroscopic properties of the bath (say 
its temperature)
 may have a non trivial influence on the resulting equations of motion,
paving the way  to the possibility of probing these quantities 
 via measurements performed on the system   alone~\cite{Brunelli2011qubit, PhysRevLett.114.220405, PhysRevA.96.012316, Campbell_2018, Kiilerich2018dynamical, PhysRevA.77.022111,Pekola,Aspelmeyer,Prokof_ev_2000,DePasquale,PhysRevA.97.012126}.

Relying on these observations, in Ref.~\cite{PhysRevA.100.042327} a 
statistics tagging scheme has been presented, allowing to 
determine the fermionic or bosonic character of a thermal bath $E$ by detecting the modifications 
induced on a quantum probing system $A$ put in thermal contact with $E$ for some proper interaction time $t$.  
The analysis was conducted assuming the temperature of the bath to be known and, most importantly, equal in the two alternative scenarios. Under this condition, waiting for the complete thermalization of $A$ (i.e. setting $t\rightarrow \infty$) 
is clearly not a valuable option to get useful information on the nature of the bath:
indeed as $t$ diverges the probe will be driven toward the same final thermal equilibrium configuration irrespectively from the statistics of $E$, hence keeping no track of its fermionic or bosonic character. As a consequence the optimal discrimination performances in Ref.~\cite{PhysRevA.100.042327} were obtained at times $t$ where the evolved state of $A$ was explicitly in a non-equilibrium condition. Superiority of non-equilibrium conditions for measurement purposes are not unique to the statistics tagging procedure discussed in~\cite{PhysRevA.100.042327}: 
a similar behaviour can be observed in thermometric tasks, when we want to infer the temperature of a bosonic bath by the same interaction with a probe. Even if the thermalized states corresponding to different temperatures can be discriminated, there is an advantage when measuring the probe at earlier times \cite{PhysRevA.91.012331, Tham:2016yu, PhysRevLett.118.130502}.  
Interestingly enough, the statistics tagging setting and the thermometric setting also share another 
common feature: indeed in both schemes the input states of the probe which ensure optimal performances correspond to energy eigenstates of its local Hamiltonian, quantum coherence playing no fundamental role in the procedure (see however \cite{Kiilerich2018dynamical}).
In an effort to check the generality of these observations (i.e. the optimality of  using non-equilibrium observation times  $t$
and energy diagonal input states of the probe), here we cast  the problem studied in Ref.~\cite{PhysRevA.100.042327} in a more complex framework by looking at the discrimination between two alternative thermal baths models which differ both in terms of their statistical properties {\it and} in terms of their associated temperatures. 
The  analysis relies on information theoretical quantities which admit clear operational interpretations in quantum metrology \cite{giovannetti2006quantum,
paris2009quantum, giovannetti2011advances, pirandola2018}.
In particular the minimization of the Helstrom probability of error \cite{helstrom1976quantum} enables us to confirm that also for the generalized statistics tagging scenario we address here,   optimal discrimination performances are obtained by 
monitoring the probe at times where it is in a non-equilibrium configuration. Yet in this case it turns out that  such result strongly relies on the possibility of exploiting coherence in the input states of $A$: indeed when restricting the study to initial configurations of the probe with no coherence among the eigenstates of the system local Hamiltonian, we can exhibit explicit examples of the model parameters for which the best discrimination conditions are only met at equilibrium.

The paper is organized as follows: in Sec.~\ref{model} we introduce the model and present the figure of merit we are going to use
in our analysis. 
Section~\ref{opt} contains the main results of the paper discussing the role
of coherent energy terms in the input state of the
probe as well as the fact that non-equilibrium detection times are not always optimal if one restrict the analysis to initial configurations which
are diagonal in the energy eigenbasis. 
In Sec.~\ref{SECIII}  the previous concepts are illustrated in a linear optical simulator~\cite{PhysRevLett.121.160602, PhysRevA.98.050101}. 
The simulation allows to mimic two different dissipative channels for a two level system~\cite{PhysRevA.95.042310}.
In the typical tagging scenario, in which we have no a priori knowledge of which one of the channels is acting on the probe, we perform a set of measurements on the system and we reconstruct the original hypothesis via suitable statistical inference.
In particular, we relied on a Bayesian technique~\cite{PhysRevA.85.043817} for constructing the error probabilities and providing a connection of this last with the theoretical figures mentioned above.
Conclusions are presented in Sec.~\ref{CONC} 
while technical material is presented in the Appendix.

\begin{figure*}
    \centering
    \vspace{.5cm}
\begin{overpic}[width=0.45\linewidth]{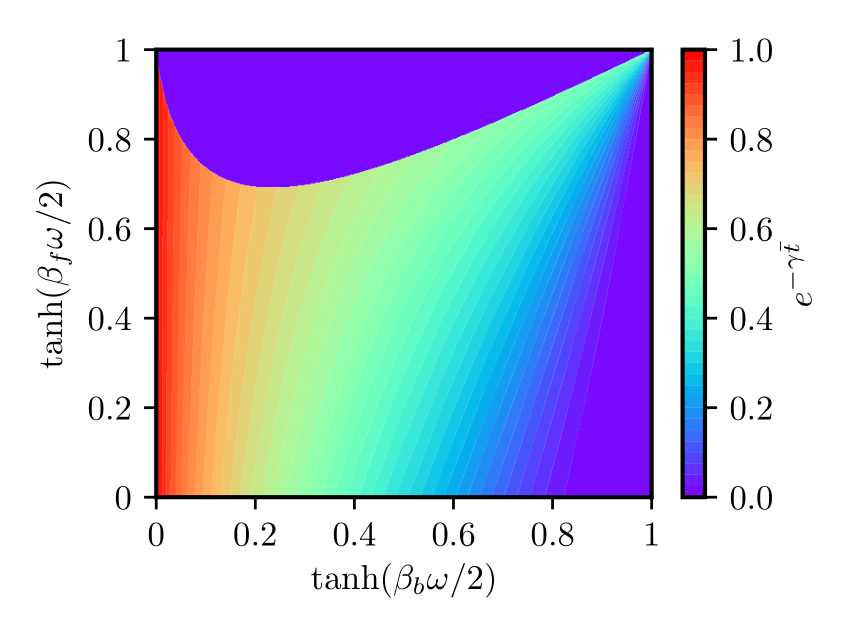}\put(30,73){(a)}\end{overpic}
\begin{overpic}[width=0.45\linewidth]{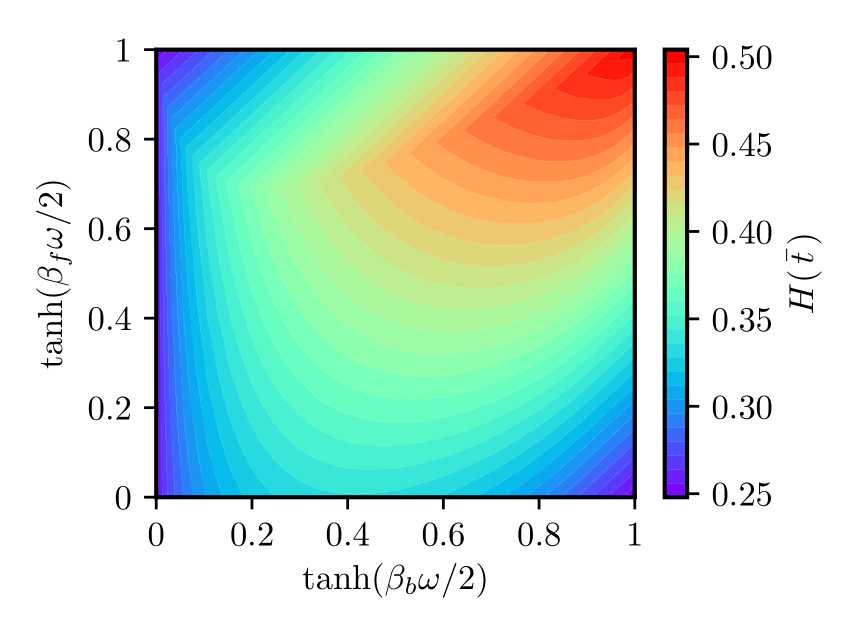}\put(30,73){}\put(-8,75){Helstrom}
\end{overpic}
\\
\vspace{.5cm}
\begin{overpic}[width=0.45\linewidth]{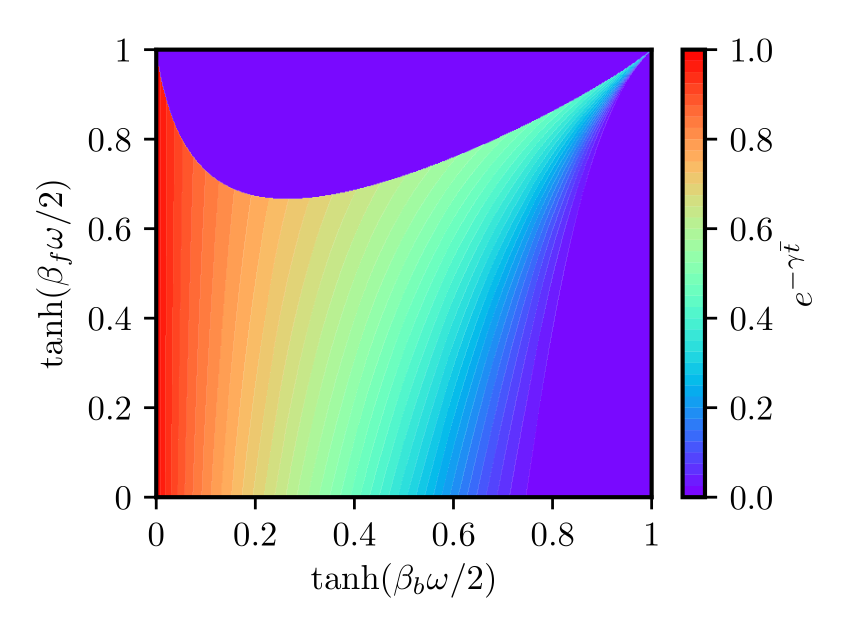}\put(30,73){(b)}\end{overpic}
\begin{overpic}[width=0.45\linewidth]{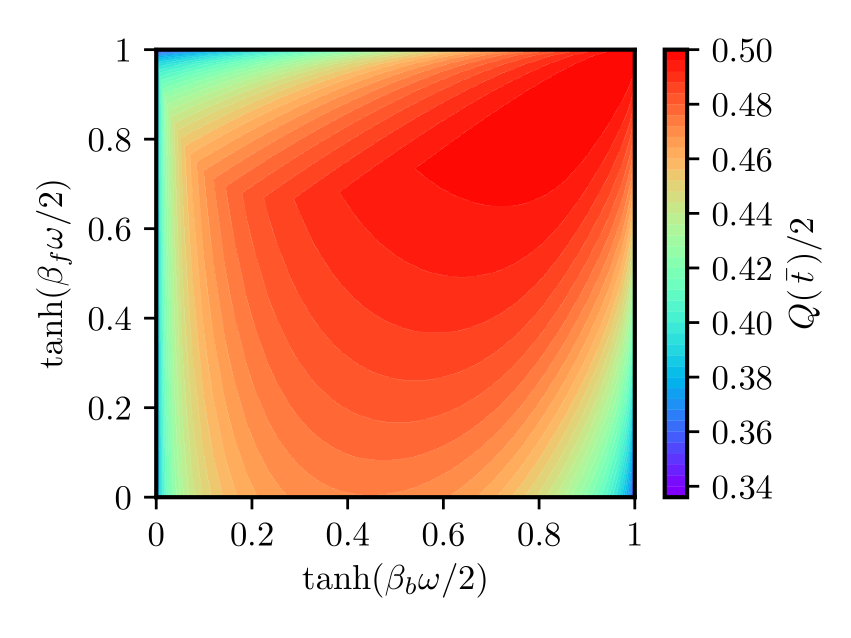}\put(30,73){}\put(-8,75){Chernoff}
\end{overpic}
    \caption{Panel (a). Left: Study of the optimal measurement time $\bar{t}$
minimizing the Helstrom error probability $H(t; 1)$ of Eq.~(\ref{MAXI}) associated to the excited input state of the probe $A$ (i.e. $a_z(0)=1$), as a function of the bosonic and fermionic bath inverse temperatures $\beta_{b}$ and $\beta_{f}$, using the convenient parametrizations indicated in the plot labels. The discontinuity in the contour plot is the boundary above which the discrimination is optimal only if performed on the steady state of the probe $(\bar{t}=\infty~,$ i.e. $e^{-\gamma \bar{t}}=0 )$, the same holding for the pathological case $\beta_{b}=\infty$ $(\tanh(\beta_{b} \omega/2)=1)$. For all the other values of the parameters $\beta_{b}$ and $\beta_{f}$ the optimal time is finite $(\bar{t}<\infty~,$ i.e. $e^{-\gamma \bar{t}}>0).$
    Right: Corresponding Helstrom probability of error $H(t; 1)$ evaluated at $t=\bar{t}$.
    Panel (b): same calculation as in (a) but using the Chernoff quantity (\ref{chernoff}) instead of the Helstrom error probability.}
    \label{fig:num1}
\end{figure*}

\section{The model}\label{model} 
The model we study 
can be schematized as follows. At time $t=0$ a two-level (qubit) quantum probe $A$ is prepared in some fiduciary initial density operator $\rho(0)$ and let interact for some time $t$ with a partially unknown environment $E$ that can be of two types: 
 bosonic at temperature $1/\beta_b$, or fermionic at temperature $1/\beta_f$ (the values $1/\beta_b$ and $1/\beta_f$ being assigned a priori). 
As in Ref.~\cite{PhysRevA.100.042327} we shall  attempt  to discriminate among the two 
alternatives by only performing measurements on the  reduced final state  $\rho(t)$  of $A$, which hence encodes all the information
about the nature of $E$ one can access. 
This allows us to describe the whole scheme as a standard hypotheses testing problem~\cite{helstrom1976quantum}  where
 one has to determine whether $\rho(t)$  corresponds to the
density matrix $\rho_b(t)$  of $A$ which one would have obtained by evolving $\rho(0)$ under the influence of the bosonic
bath of temperature $1/\beta_b$, or to $\rho_f(t)$, which instead one would have obtained by evolving the same
$\rho(0)$ under the influence of the fermionic 
bath of temperature $1/\beta_f$.
To quantify our ability in  discriminating 
between these scenarios 
we can then use the Helstrom error probability (HEP) functional 
\begin{eqnarray} 
H(\rho_{b}(t), \rho_{f}(t)) &:=& \frac{1}{2}  - {\frac{1}{4}}\| \rho_b(t)-\rho_f(t)\|_1 \;, 
   \label{eq:hel} 
\end{eqnarray}
with $\|\cdots \|_1$ being the trace-norm symbol.
This quantity, bounded between $[0,1/2]$, provides
the smallest probability of error one can get by optimizing over all possible  
measurements performed on a single copy of $\rho(t)$~\cite{helstrom1976quantum}:
accordingly, having $H(\rho_{b}(t), \rho_{f}(t))=0$ corresponds to perfect distinguishable 
configurations, while having $H(\rho_{b}(t), \rho_{f}(t))=1/2$ corresponds to absolutely indistinguishable
configurations.

In order to get an analytical expression for (\ref{eq:hel}) we 
assign $\rho_b(t)$ and $\rho_f(t)$ in terms 
 two independent Gorini-Kossakowsky-Sudarshan-Lindblad 
 master equations for $A$ obtained
 under standard 
 weak coupling system-bath assumptions~\cite{lindblad1976generators, gorini1976completely}.
Moving into the interaction picture representation we write them as~\cite{PhysRevA.100.042327, farina2019open, Esposito2010}
\begin{equation}
\dot{\rho}_q(t)=\gamma[1+s_q \mathcal{N}_q(\beta_q)] \mathcal{D}_{\sigma_-}[{\rho}_q(t)]
+ \gamma \mathcal{N}_q(\beta_q) \mathcal{D}_{\sigma_+}[{\rho}_q(t)]~,
\label{master-equation}
\end{equation}
the index
$q=f,b$ 
referring to the two hypothetical initial configurations of the bath.
In the above expression $s_q=+(-)1$ for $q=b ~(f)$, 
 $\gamma$ is the inverse time constant associated to each elementary excitation/de-excitation process, 
\begin{eqnarray}
\mathcal{D}_{\sigma_\pm}[\cdots]:=\sigma_{\pm} [\cdots] \sigma_{\pm}^\dagger - \frac{\sigma_{\pm}^\dagger \sigma_{\pm}
[\cdots] +  [\cdots]  \sigma_{\pm}^\dagger \sigma_{\pm}}{2},  %
\end{eqnarray}
represent the Lindblad dissipators associated respectively with the system ladder operators 
$\sigma_-=\ket{0}\bra{1}$ and $\sigma_+=\ket{1}\bra{0}$ ($|0\rangle$ and $|1\rangle$ representing respectively the ground and excited state of $A$), while finally 
\begin{eqnarray}
\mathcal{N}_q(\beta_q):=\frac{1}{e^{\beta_q \omega}-s_q} \;,\label{eq:N}
\end{eqnarray}
is the Bose-Einstein (Fermi-Dirac) distribution for $q=b ~ (f)$, with $\omega$ being an effective energy parameter \cite{Esposito2010,petruccione} that contains a contribution from the bare energy of $A$ and from 
the chemical potential of the baths \footnote{We suppose $\omega$ to be the same for $b$ and $f$. When the chemical potential is different between the fermionic and bosonic cases we can opportunely redefine $\beta_f$ and $\beta_b$ to preserve the Eq. (\ref{eq:N})}.
Introducing the Pauli vector operator $\vec{\sigma} : = ( \sigma_x,\sigma_y,\sigma_z)$,
and writing the  density matrix of the system in the Bloch vector formalism
$\rho_{q}(t) = \frac{1 + \vec{\sigma} \cdot \vec{a}^{(q)}(t)}{2}$,  
Eq.~(\ref{master-equation}) can then be conveniently casted in the form 
 \begin{eqnarray}   
 \dot{a}^{(q)}_z(t) &=& -\gamma_{q} a_z^{(q)}(t) - \xi_q,  \label{eq:dyn} \\
 \dot{a}^{(q)}_{x,y}(t)&=& - \frac{ \gamma_{q} }{2} a_{x,y}^{(q)}(t) ,
 \nonumber
 \end{eqnarray} 
where now 
\begin{eqnarray} \begin{array}{lll} 
\gamma_{b} := \gamma \coth(\beta_{b} \omega/2),& & \gamma_{f} := \gamma\;, \\ 
\xi_b :=\gamma, & & \xi_{f} : = \gamma \tanh(\beta_{f} \omega/2)\;, 
\end{array} \end{eqnarray}  
showing that  in the case of equal temperatures, the evolution occurs at faster scales for the bosonic bath scenario.
Explicit integration of (\ref{eq:dyn}) leads finally 
 to the solution
 \begin{eqnarray} \label{soldyn}
a_z^{(q)}(t) &=& e^{-\gamma_q t}( a_z(0) -a_z^{(q)}(\infty)) + a_z^{(q)}(\infty) \;, \\  
a_{x,y}^{(q)}(t) &=& e^{-\gamma_q t/2} a_{x,y}(0) , \nonumber 
\end{eqnarray} 
with $a_{x,y,z}(0)$ being the cartesian components of the Bloch vector associated with the input state $\rho(0)$ of $A$, while
\begin{eqnarray} 
a_z^{(q)}(\infty)= - \tanh(\beta_{q} \omega/2)\;,
\end{eqnarray} 
defining the equilibrium (thermal) configuration of the system (of course $a_{x,y}^{(q)}(\infty)= 0$).

 \section{Discrimination perfomances} \label{opt}

Using the fact that the trace-norm of the difference between $\rho_b(t)$ and $\rho_f(t)$  is just  given by the
Cartesian distance
  $|\vec{a}_b(t)-\vec{a}_f(t)|$  of the associated  3D Bloch vectors, 
from (\ref{soldyn}) it follows that Eq.~(\ref{eq:hel}) can be expressed as  
\begin{widetext} 
\begin{eqnarray}  \nonumber
H(\rho_{b}(t), \rho_{f}(t))  = \frac{1}{2} - \frac{1}{4}&& 
  \Big\{\Big[(e^{-\gamma_f t}- e^{-\gamma_b t}) {a}_z(0) + a_z^{(f)}(\infty) (1- e^{-\gamma_f t})  -  a_z^{(b)}(\infty)  (1- e^{-\gamma_b t})  \Big]^2 \\
&&+ \label{QUESTA1}
(e^{-\gamma_f t/2}- e^{-\gamma_b t/2})^2 \big(| \vec{a}(0)|^2 - {a}^2_z(0)   \big) \Big\}^{1/2}\;.
\end{eqnarray} 
A close inspection reveals that all pure input states $\rho(0)$ with the same initial value of $a_z(0)$ achieve the same performance
(this simply follows from  the symmetry of Eq. (\ref{eq:dyn}) around the $z$-axis). 
Furthermore for all assigned values of $t$ and ${a}_z(0)$, one may notice that  the associated HEP can be reduced by setting 
the length of $\vec{a}(0)$ at its maximum $1$, i.e. imposing the initial state of the probe to be pure. This leads to 
\begin{eqnarray}  \nonumber
H(\rho_{b}(t), \rho_{f}(t))\Big|_{\text{pure}} = H(t; a_z(0)) :  = \frac{1}{2} - \frac{1}{4}&& 
  \Big\{\Big[(e^{-\gamma_f t}- e^{-\gamma_b t}) {a}_z(0) + a_z^{(f)}(\infty) (1- e^{-\gamma_f t})  -  a_z^{(b)}(\infty)  (1- e^{-\gamma_b t})  \Big]^2 \\
&&+ \label{QUESTA}
(e^{-\gamma_f t/2}- e^{-\gamma_b t/2})^2 \big(1 - {a}^2_z(0)   \big) \Big\}^{1/2}\;,
\end{eqnarray} 
\end{widetext} 
which only depends on the $z$-component  ${a}_z(0)\in [-1,1]$ of the unit vector $\vec{a}(0)$. 
It is worth recalling that  fixing $a_z(0)=1$ ($a_z(0)=-1$) corresponds to initialize $A$ into the excited  state $|1\rangle$ (ground state $|0\rangle$) of its local Hamiltonian. On the contrary, in the pure case scenario we are facing in Eq.~(\ref{QUESTA}),  the condition
$|a_z(0)|<1$ identifies input states of the probe which are proper superpositions of the energy eigenstates of the model.
Our next goal is to  minimize  $H(t;a_z(0))$ with respect to all possible choices of  ${a}_z(0)$ and of the  evolution time $t$,
for given values of the temperatures $1/\beta_f$ and $1/\beta_b$. 
Before doing so, however, we find useful to consider first 
what happens when $a_z(0) =1$, 
a choice that is known to provide the best discriminating strength for statistical tagging under equal bath temperature assumption (i.e. $\beta_f = \beta_b$)~\cite{PhysRevA.100.042327}  and for thermometry~\cite{PhysRevA.91.012331}.

\subsection{Input excited state} \label{INPUTEX}
Setting $a_z(0) =1$, i.e. assuming $A$  to be initialized in the excited state $|1\rangle$  of the model, Eq.~(\ref{QUESTA}) reduces to 
\begin{widetext}
\begin{eqnarray}  
 H(t; 1)  = \frac{1}{2} - \frac{1}{4}
\Big|e^{-\gamma_f t}- e^{-\gamma_b t} + a_z^{(f)}(\infty) (1- e^{-\gamma_f t})  -  a_z^{(b)}(\infty)  (1- e^{-\gamma_b t}) 
 \Big|\;,\label{MAXI} 
\end{eqnarray} 
\end{widetext} 
which we minimize numerically with respect to $t$  as a function of $\beta_f$ and $\beta_b$. 
The optimal times $\bar{t}$ we obtain and the corresponding values of $H(\bar{t};1)$ are reported in Fig.~\ref{fig:num1}(a) 
(left and right plots, respectively). The plot reveals an {asymmetry}: 
for $\beta_b\geq \beta_f$ (fermion bath hotter than bosonic bath) the best discrimination is still attained at finite time ($\bar{t}<\infty$) where $A$ has not achieved full thermalization and is hence
in a non-equilibrium configuration in line with the findings of Ref.~\cite{PhysRevA.100.042327}; on the contrary for $\beta_b<\beta_f$
(fermion bath cooler than bosonic bath)
 it can be more convenient to discriminate the two channels by exploiting the steady state properties ($\bar{t}=\infty$). This happens above the {critical curve} that defines the discontinuity in the left contour plot of Fig.~\ref{fig:num1}(a).
An analytical treatment of this transition is given in Appendix~\ref{appendix-excited}, from which 
it results that expressed in the $x=\tanh(\beta_f \omega/2)$, $y=\tanh(\beta_b \omega/2)$ coordinates of Fig.~\ref{fig:num1}, 
 such critical curve is identified by solving the following set of transcendental equations
\begin{eqnarray}   \left\{ \begin{array}{l} (2- e^{-\tau})(1+x) - (2- e^{- \frac{\tau}{y}})(1+y) =0, \\\\
        e^{-\tau} (1+x) - e^{- \frac{\tau}{y}} y^{-1} (1+ y)= 0,
        \end{array} \right.
        \label{critical-curve2}\end{eqnarray} 
        with $\tau\geq0$. 
We remark that the core of the above observation remains unchanged when we 
evaluate the discrimination efficiency of the process
adopting different 
figures of merit. 
For instance in Fig~\ref{fig:num1}(b)  we focus on  the 
Chernoff quantity~\cite{chernoff-prl-2007, chernoff-pra-2008}
\begin{equation}
Q(\rho_{b}(t), \rho_{f}(t)): = {\rm min}_{r\in [0,1]} ~\mbox{Tr}[ \rho_b^r(t) \rho_f^{1-r}(t)]\;, 
\label{chernoff}
\end{equation}
which via the inequality 
\bg  H(\rho^{\otimes N}_{b}(t), \rho^{\otimes N}_{f}(t)) \leq \frac{Q(\rho_{b}(t), \rho_{f}(t))^N}{2},  \eg
gives a bound to the asymptotic rate of HEP computed in the  case when 
 one has the possibility of extracting information from  $N$ identical copies of the final  state of~$A$.
The optimal values  of $\bar{t}$ 
obtained by numerically minimizing~(\ref{chernoff}) when initializing $A$ in the excited state $|1\rangle$, are presented
in Fig~\ref{fig:num1}(b) exhibiting a critical trade-off analogous to the one observed  in Fig~\ref{fig:num1}(a):
if we restrict the analysis to the case where $A$ is set into the excited state there are configurations 
of the model  where the optimal discrimination efficiency
is attained only letting the system to reach its equilibrium configuration.

\begin{figure*}
    \centering
    \vspace{.5cm}
\begin{overpic}[width=0.45\linewidth]{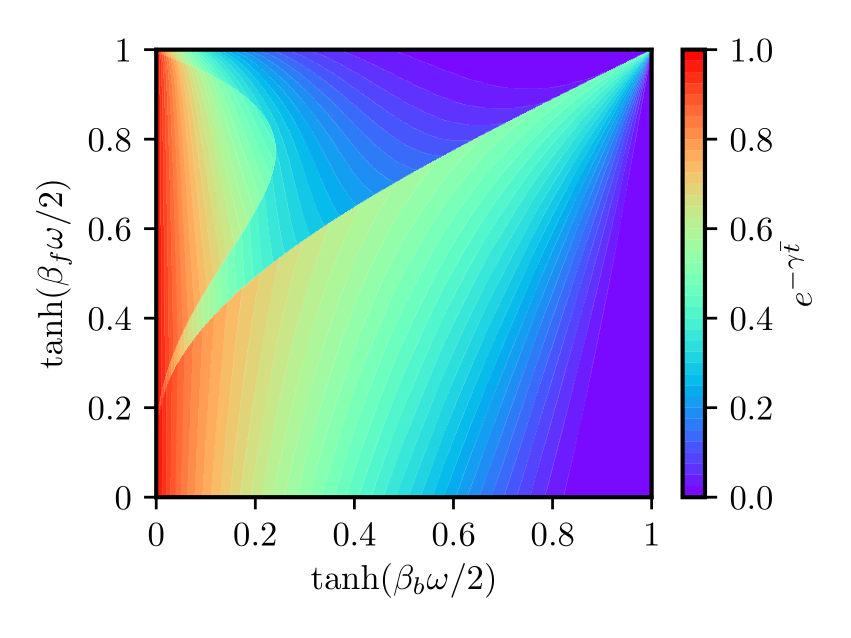}\put(30,73){(a)}\end{overpic}
\begin{overpic}[width=0.45\linewidth]{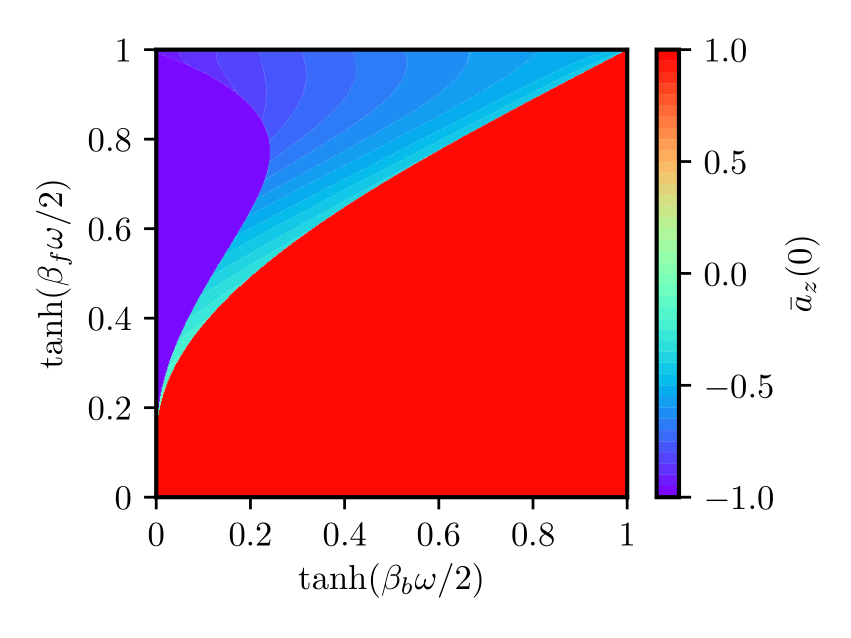}\put(30,73){(b)}
\end{overpic}
\\
\vspace{.5cm}
\begin{overpic}[width=0.45\linewidth]{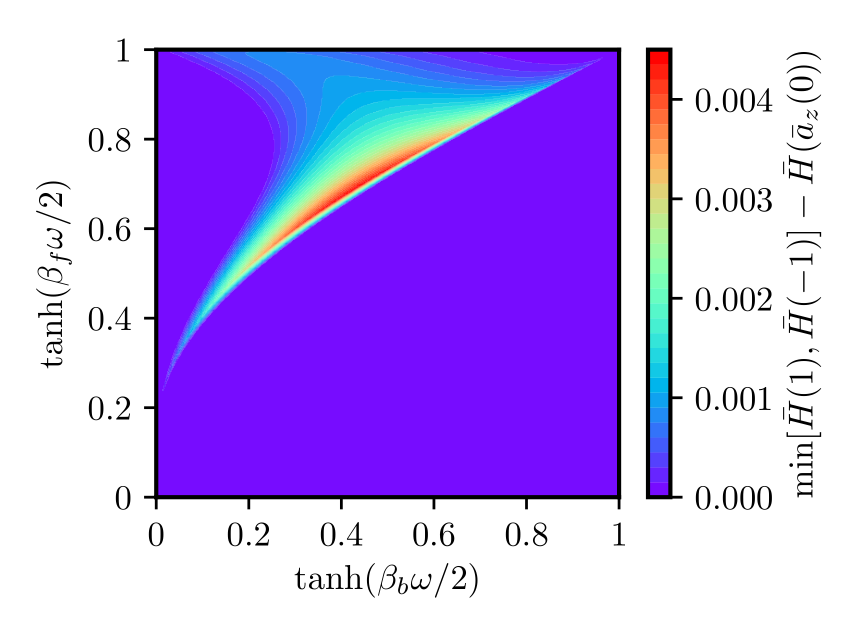}\put(30,73){(c)}\end{overpic}
\begin{overpic}[width=0.45\linewidth]{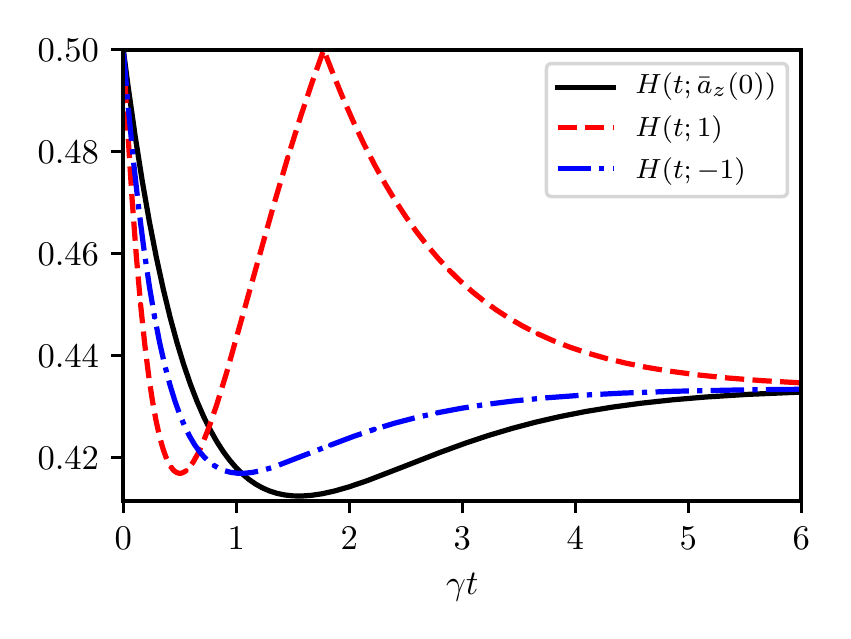}\put(30,73){(d)}
\end{overpic}
\caption{
Panels (a)-(c): Optimization of the HEP $H(t; {a}_{z}(0))$ of Eq.~(\ref{QUESTA}) both over time and over the input state of the probe, the last being a generic pure state with a certain value $a_z(0)$ of the z-component of the Bloch vector. We report the following contour plots with respect to the bath inverse temperatures $\beta_f$ and $\beta_b~.$ 
The minimum $\bar{H}(\bar{a}_{z}(0))$ of the HEP is achieved at a certain time $\bar{t}$ (a) and
for an optimal value $\bar{a}_z(0)$ of $a_z(0)$ (b). 
The advantage coming from allowing coherent superpositions is presented in (c), where we show the gap between the (generally overestimated) quantity obtained by restricting the analysis only to $a_z(0)\in \{1,-1\}$ and the optimal value $\bar{H}(\bar{a}_{z}(0))~.$
Notice that for $\beta_f~,$ $\beta_b$ and $\bar{t}$ we used the convenient parametrizations indicated in the plot labels. 
Panel (d):  dynamical evolution of $H(t; {a}_{z}(0))$ for a case 
($\tanh(\beta_f \omega/2)\approx 0.68~,$
$\tanh(\beta_b \omega/2)\approx 0.41$)
in which coherent superpositions ($\bar{a}_z(0)\approx -0.42$) 
give better performances than the energy eigenstates ($a_z(0)\in \{1,-1\}$) as input of the probe. Notice that at 
$t\approx 1.8\gamma^{-1}$ the HEP associated with excited state reaches the worst case value $1/2$ indicating zero
susceptibility of the probe. 
}
\label{fig:z0}
\end{figure*}

\subsection{Optimal input states of the probe}

In this section we now exploit the full domain of possibilities offered by 
the model, minimizing the HEP value~(\ref{QUESTA}) not just with respect to $t$, but also  with respect to the full domain of   $a_z(0)$, hence
including the possibility of using input states of $A$ which explicitly exhibit coherence superpositions among the excited and ground state of the model. An indication that  such special states could be of some help in improving the performance of the scheme follows by observing that
for $|a_z(0)|<1$ it is not possible to find  times $t>0$ such that $H(t; a_z(0))$ reaches the worst case value of $1/2$
corresponding to an absolute impossibility of  distinguishing among the two bath scenarios. 
This implies that coherent energy input states ensure a non-trivial susceptibility of the probe for all choices of $t$, something 
that, on the contrary, is not generally granted by setting $a_z(0)=\pm 1$ which, as discussed in 
Appendix \ref{app:generic1}, allows for 
 crossing points between the trajectories $\rho_b(t)$ and $\rho_f(t)$. 
Values of  $|a_z(0)|<1$ can however  do much more than this  and in some  regimes, they 
also give the absolute best performance we can aim to:
 the details of the analysis are provided in Appendix \ref{app:generic2} while 
in Fig.~\ref{fig:z0} we illustrate 
the optimization of the HEP $H(t; {a}_{z}(0))$ over time and input state of the probe,
 as a function of the bath inverse temperatures $\beta_f$ and $\beta_b$.

The first thing to be noticed
  is that now, at variance with the input excited state case 
 discussed in Sec.~\ref{INPUTEX},  
 the optimal times $\bar{t}$ are always finite 
 apart from the asymptotic regimes where the bosonic temperature converges to zero
 (i.e. $\beta_b\rightarrow \infty$) 
 --  compare  Fig.~\ref{fig:z0} (a) with   
 the left plot of
 Fig.~\ref{fig:num1} (a). 
 This shows that optimality of non-equilibrium probing times is fully restored once we do not restrict the  probe 
 input state to specific conditions. 
 Secondly Fig. \ref{fig:z0} (b) reveals that, while using energy eigenstates (either excited or ground states) 
 of the probe as input is optimal for most of the choices
 of the system parameter setting, there is a non trivial temperatures regime in which a coherent ($|a_z(0)| <1$)
  initial preparation is fundamental to reach the best performance.
 More specifically, there is numerical evidence that whenever the fermionic bath is hotter than the bosonic one ($\beta_b\geq \beta_f$), choosing the 
  excited state of $A$ as input is still the right choice to provide optimal discrimination performances. The situation changes however 
 if the fermionic bath is cooler  than the bosonic one ($\beta_b <  \beta_f$): here
 the optimal input choice depends on the specific values of the temperatures and in particular for sufficiently large  $\beta_f$
 coherent energy states can dominate (notice also that, for small values of $\beta_b$, the optimal input can be the ground state of $A$). 
   These facts are also enlightened in Fig. \ref{fig:z0} (c) in which we show the gap between the minimum of $H(t; a_z(0))$
 obtained by restricting the optimization only to $a_z(0) = 1$ and $a_z(0) =-1$ and the optimal value $\bar{H}(\bar{a}_{z}(0))$ obtained by allowing also energy coherent preparations. 

In panel (d) of Fig. \ref{fig:z0} we finally present as an example the temporal  evolution of the HEP for a specific choice of the temperatures 
that admits as optimal the value $\bar{a}_z(0)\approx -0.42$ that identifies a 
coherent superposition of energy eigenstates.
In such a plot we show $H(t; \bar{a}_z(0))$ aside with the HEP values $H(t; -1)$ and $H(t; 1)$ associated with the ground and excited input state of $A$. 
Notice that while for small $t$, $H(t; -1)$ and $H(t; 1)$ perform better than $H(t; \bar{a}_z(0))$, in the long run the latter gives the lowest HEP values and
 leads to the identification of  the optimal time as $\bar{t} \approx 1.6 \gamma^{-1}$ --  see Appendix for more on this.
 Notice also that  at $t\approx 1.8\gamma^{-1}$, we have  $H(t; 1)=1/2$ indicating that at this special
 time the probe intialized into the excited state looses all its ability in discriminating between the two alternative hypothesis: 
 on the contrary, as anticipated in the introductory paragraphs of the section,  $H(t; \bar{a}_z(0))$ remains strictly below the 
 $1/2$ value for all positive $t$.

\begin{figure}[h!]
\includegraphics[width=\columnwidth]{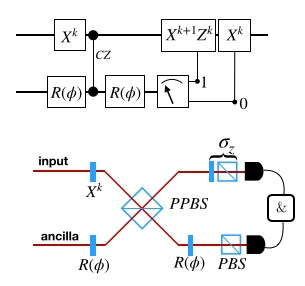}
\caption{Linear optical simulator. The state at time $\tau$ can be simulated by mixing with weights $w_k$ the actions of two channels ($k=0$ and $k=1$), associated to excitation and de-excitation processes, for a given rotation $R(\phi)$ (upper panel). By tuning $w_k$ and $\phi$ evolutions at different temperatures and at variable times are simulated~\cite{PhysRevLett.121.160602,PhysRevA.98.050101}. 
This is implemented in a linear-optical setup, based on polarisation coding on two photons from a down-conversion source (lower panel). Single-qubit operations are either implemented by half-wave plates, or in post-processing of the data. The input state is fixed in the horizontal polarisation. Due to the use of a single partially polarising beam splitter (PPBS), there is a different transmission probability for the horizontal and vertical components, which is compensated by biassing the second $R(\phi)$ rotation~\cite{PhysRevLett.99.250505}. Further, the weights $w_k$ have to be modified accordingly.}
\label{fig:setup}
\end{figure}

\section{Discrimination experiment in an optical simulator}\label{SECIII} 

We can illustrate these concepts in a simulated thermalisation, carried out with a pair of qubits; the necessary gate is implemented by means of optical elements and coincidences counts. The setup, illustrated in Fig.~\ref{fig:setup}, follows closely our previous work in Ref.~\cite{PhysRevLett.121.160602}. We stress that our simulator cannot replicate directly the bosonic/fermionic nature of the bath; the control parameters are exclusively the decay rates $\gamma_f$ or $\gamma_b$ in Eq. (\ref{eq:dyn}), and the population of the final thermal state. In this respect, our implementation is a synthesis of the output state. Therefore, we focus on the information content of the probe, rather than the interaction process.

\begin{figure*}[t]
\includegraphics[width=\textwidth]{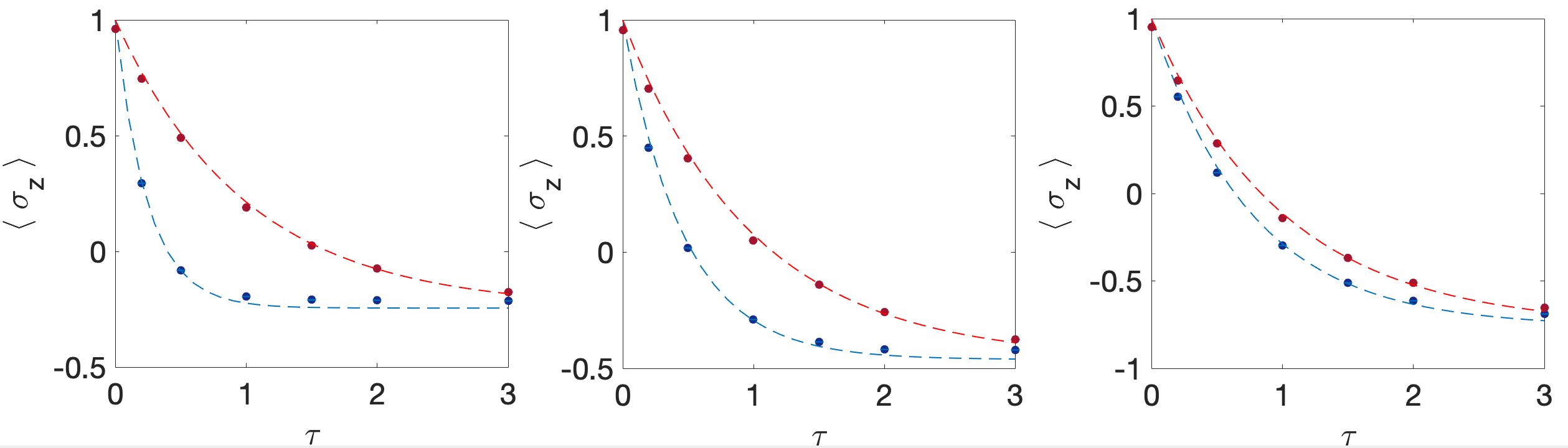}
\caption{Simulated thermalization dynamics of the probe initialized in the excited state. The behaviour of the expectation value $\langle \sigma_z \rangle$ as a function of the time $\tau=\gamma t$ is reported for ($\omega\equiv 1$) $\beta_b=\beta_f=0.5, 1, 2$ (left, middle, and right panel, respectively). Blue (red) experimental points refer to the bosonic (fermionic) statistics of the bath. Error bars are smaller than the size of the points.}
\label{fig:sigma}
\end{figure*}

We consider a two level system initialized in the excited state as the input probe. The expectation values of $\sigma_z$ measured as a function of the normalized time $\tau=\gamma t$ are shown in Fig.~\ref{fig:sigma} for different inverse temperatures $\beta \omega=0.5, 1, 2,$ taken equal for fermionic and bosonic baths. 
The two different curves in each panel illustrate how the decay rate of the probe state gets modified by the two different statistics.

In a discrimination experiment, the sought outcome is a binary decision on which one of the two hypotheses gives a closer description of the data~\cite{PhysRevLett.97.193601,PhysRevLett.94.220406, PhysRevLett.118.030502, Bina:17,DiMario:18, Becerra}. These will be obtained as outcomes of a suitable observable, selected according to the initial state and the measurement time. For our choice of initial state, this observable always coincides with $\sigma_z$.
In many different (and independent) runs of the experiment,  one collects $N_0$ events for the eigenvalue -1 and $N_1$ events for the eigenvalue +1 of $\sigma_z$. Since the probabilities of obtaining either result on a single copy are $P_i= (1+(-1)^{i+1} \langle \sigma_z \rangle^{(q)}(t) )/2$, where the value of $\langle \sigma \rangle_z^{(q)}$ is the expectation value predicted by the experiment, the composite probability is $\mathbf{P}=P_0^{N_0}P_1^{N_1}$.


Clearly, the probability $\mathbf{P}$ depends on the bath statistics and temperature through the expectation value $\langle \sigma_z \rangle$. We can thus interpret $\mathbf{P}$ as a conditioned probability $\mathbf{P}(N_0,N_1\vert X)$ of the whole experimental run, given the condition $X$ of the bath. Invoking Bayes theorem this writes: 
\begin{equation}
    \mathbf{P}(X\vert N_0, N_1) = \frac{1}{\mathcal{N}}\mathbf{P}(N_0,N_1\vert X)P(X),
\end{equation}
where $\mathcal{N}$ is a normalization constant and $P(X)$ is the {\it a priori} probability which we take to be flat $P(b)=P(f)=1/2$. The decision criterion is that when $P(b\vert N_0, N_1)>P(f\vert N_0, N_1)$, the bath is identified as bosonic with inverse temperature $ \beta_b$, otherwise as fermionic with inverse temperature $\beta_f$.

\begin{figure*}[h!]
\includegraphics[width=\textwidth]{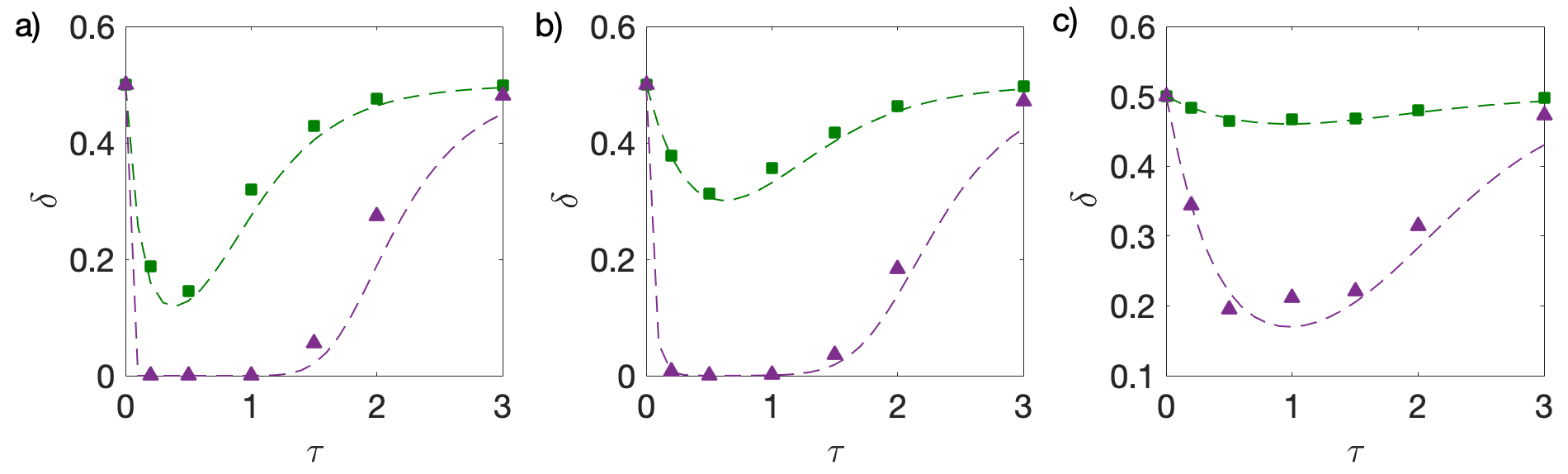}
\caption{Estimation errors for statistical tagging. The error $\delta$ is reported for a) $\beta_b=\beta_f=0.5$ b) $\beta_b=\beta_f=1$ c) $\beta_b=\beta_f=2$ ($\omega\equiv 1$). In all panels, points are evaluations of the expected errors based on the experimental probabilities, and the curves are the ideal cases for $N=10$ (green squares) or $N=100$ (purple triangles). The probe was initialized in the excited state.}
\label{fig:error}
\end{figure*}

In accordance with the literature \cite{helstrom1976quantum, nielsen2002quantum}, we 
quantify the expected discrimination error as:
\begin{equation}
\label{delta-error}
 \delta = \frac{1}{2}\left( P(b,\beta_b \vert N_{0}^{f,\beta_f}, N_{1}^{f,\beta_f}) + P(f,\beta_f\vert N_{0}^{b,\beta_b}, N_{1}^{b,\beta_b}) \right),
\end{equation}
where we have fixed $N_{i}^{q,\beta_q}=P_i \cdot N$, ($N$=10,100) as the ideal limit.
The first case we analyse is that of statistical tagging $\beta_b=\beta_f$, for which optimal discrimination necessarily occurs at finite times. 
In Fig. \ref{fig:error} we show the behaviour of $\delta$ for $\beta_b=\beta_f$. 
We notice that the small discrepancies observed with respect to the theory do not affect the estimation significantly. It appears evident how, for high temperatures, the choice of a preferable discrimination time becomes less strict with increasing copies $N$. On the other hand, the proximity of the two curves in Fig.~\ref{fig:sigma}(c) is reflected in the fact that at low temperatures more copies are needed for a fully reliable discrimination.

\begin{figure*}[h!]
\includegraphics[width=\textwidth]{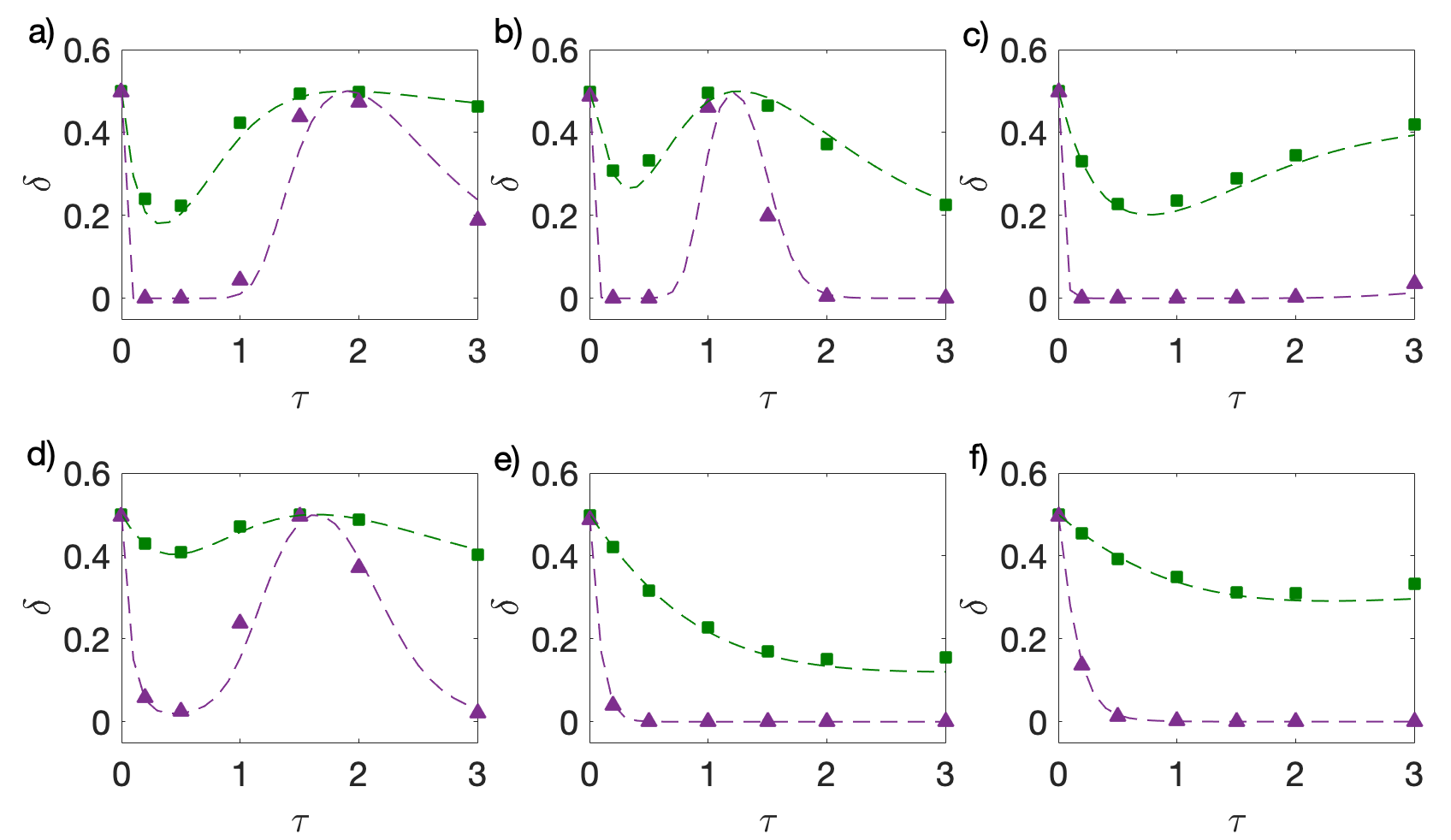}
\caption{Estimation errors for general bath discrimination. The error $\delta$ is reported for a) $\beta_b=0.5$, $\beta_f=1$  b) $\beta_b=0.5$, $\beta_f=2$ c) $\beta_b=1$, $\beta_f=0.5$ d) $\beta_b=1$, $\beta_f=2$  e)$\beta_b=2$, $\beta_f=0.5$ f) $\beta_b=2$, $\beta_f=1$ ($\omega\equiv 1$). In all panels, points are evaluations of the expected errors based on the experimental probabilities, and the curves are the ideal cases for $N=10$ (green squares) or $N=100$ (purple triangles).
The probe was initialized in the excited state.}
\label{fig:horror}
\end{figure*}

\begin{figure*}[h!]
    \centering
    \vspace{.5cm}
\begin{overpic}[width=0.45\linewidth]{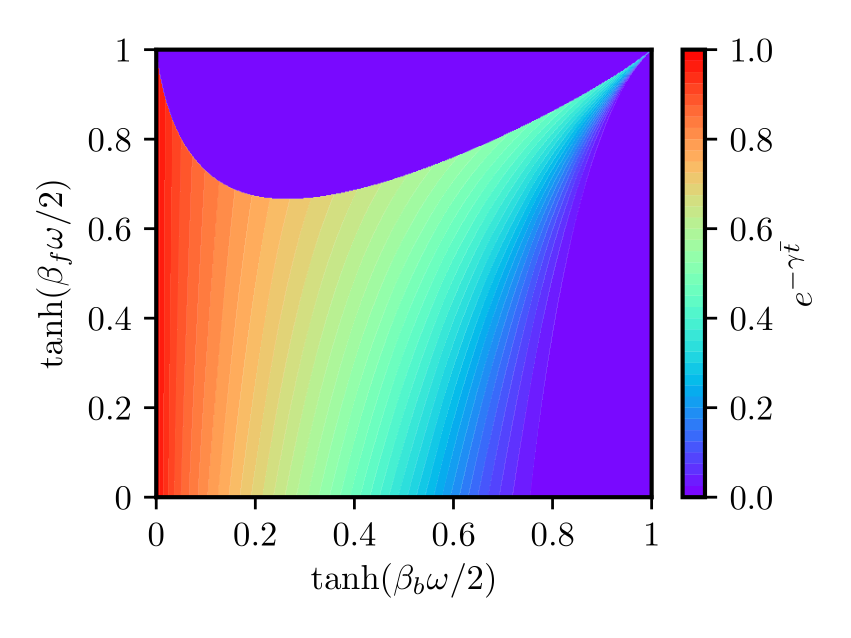}\put(30,73){(a)}\end{overpic}
\begin{overpic}[width=0.45\linewidth]{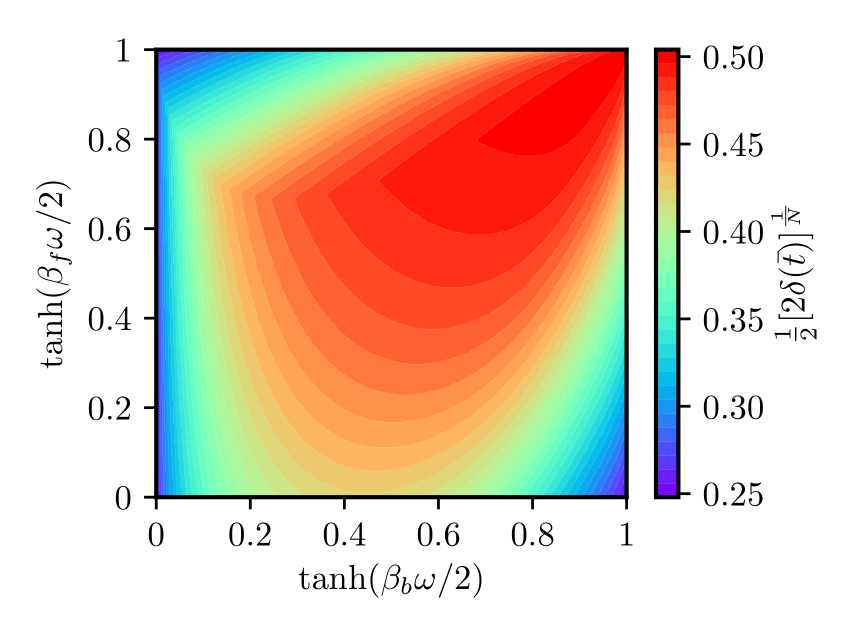}\put(30,73){}\put(-8,75){Bayes (N=10)}
\end{overpic}
\\
\vspace{.5cm}
\begin{overpic}[width=0.45\linewidth]{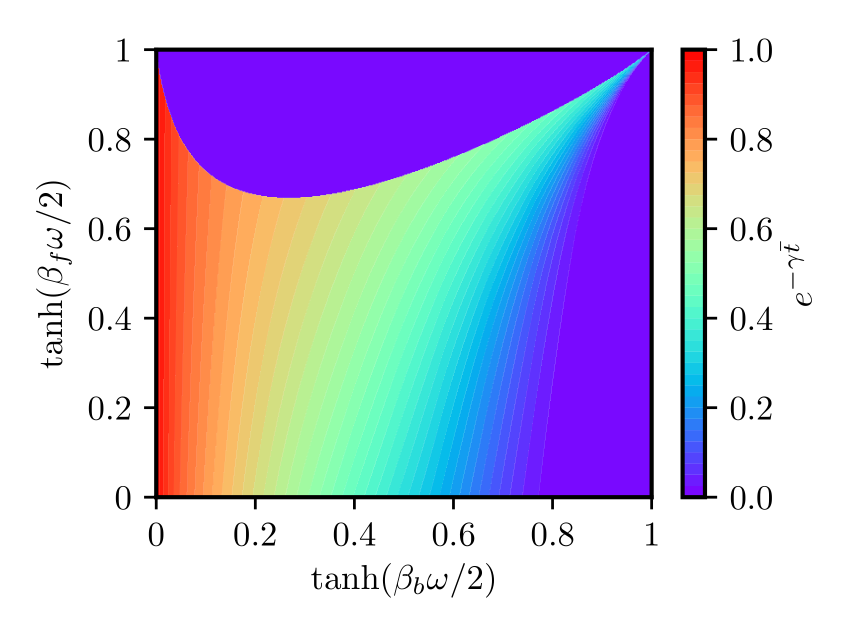}\put(30,73){(b)}\end{overpic}
\begin{overpic}[width=0.45\linewidth]{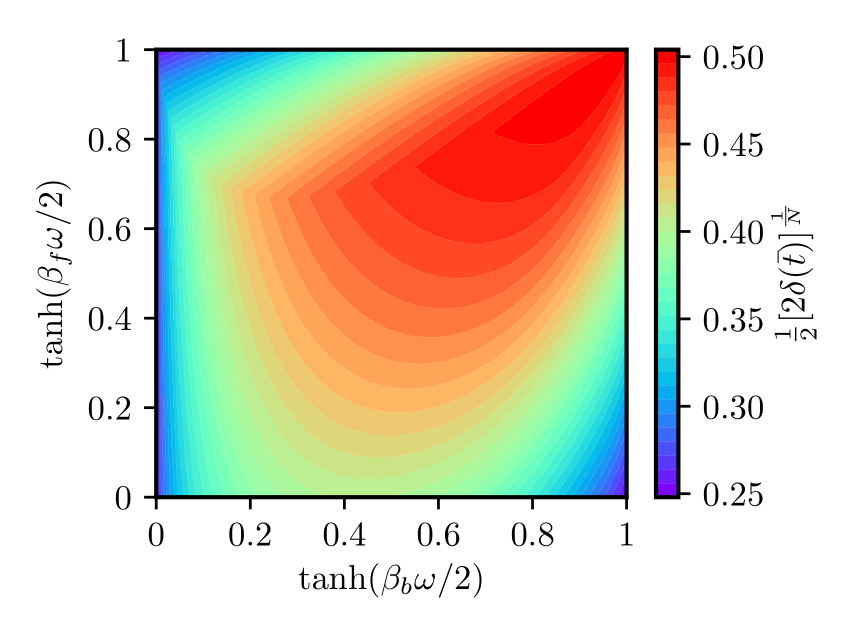}\put(30,73){}\put(-8,75){Bayes (N=100)}
\end{overpic}
    \caption{Same study as in Fig.~\ref{fig:num1} but for the error probability $\delta$ of Eq.~(\ref{delta-error}) based on the Bayesian approach for N=10 (Panel a) and N=100 (Panel b). To allow a direct comparison for different N and with the figures of merit of Fig.~\ref{fig:num1}, we report instead of $\delta$ its rescaled version $1/2 [2 \delta]^{1/N}.$ We remark that as for Fig.~\ref{fig:num1}  the probe was initialized in the excited state.}
    \label{fig:bayes-bar}
\end{figure*}

Concerning the more general scenario of different temperatures and statistics, we have evaluated $\delta$ for all permutations of $\beta \omega= 0.5,1, 2$ and for $N=10, 100$ in the same ideal limit as above. 
The results are shown in Fig. \ref{fig:horror}. 
Notably, for $\beta_b<\beta_f$ there is a special time instant where the discrimination is impossible, in analogy to what we obtained for the HEP 
- see Panel (d) of Fig.~\ref{fig:z0}. This can be observed in Panels (a), (b) and (d) of Fig.~\ref{fig:horror} where, contrarily to the other Panels, $\delta$ takes the value $1/2$ at an intermediate time. We report in Fig.~\ref{fig:bayes-bar} contour plots showing the calculation of the optimal measurement time and of the corresponding minimized error probability when using the Bayesian method in the ideal situation. Consistently, the results mimic the ones obtained via Helstrom and Chernoff approaches - see Fig.~\ref{fig:num1}.

{\color{magenta}
}

\begin{figure*}[h!]
\includegraphics[width=\textwidth]{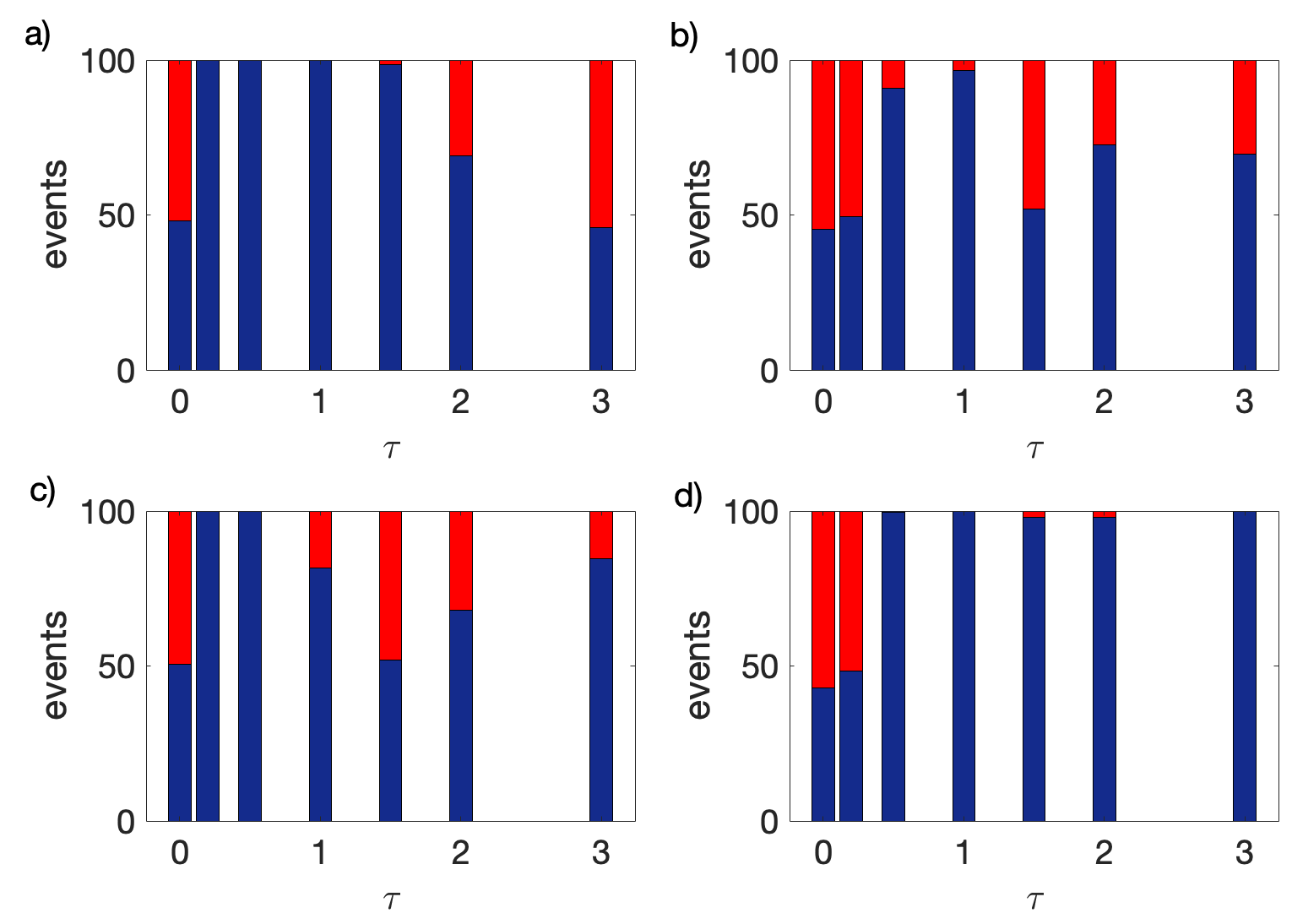}
\caption{Bayesian bath discrimination. The histogram report the events correctly identified in blue (lower part of the histogram bars), and the incorrect ones in red (upper part of the bar). The discrimination tasks are a) statistical tagging with $\beta_b=\beta_f=0.5$, b)statistical tagging with $\beta_b=\beta_f=2$, c) bath discrimination with $\beta_b=0.5$, $\beta_f=1$, c) bath discrimination with $\beta_b=2$, $\beta_f=1$ ($\omega\equiv 1$). The probe was initialized in the excited state.}
\label{fig:probs}
\end{figure*}

 We then carry out the actual discrimination protocol as follows. We generate, based on the experimental values of $P_0$ and $P_1$ a vector of $N=100$ outcomes $0,1$~\footnote{This is achieved by generating a random number $r$ uniformly between 0 and 1; if $r<P_0$, then $N_0$ is incremented by one unit (starting from $N_0=N_1=0$), otherwise $N_1$ is incremented.}.  This is a reliable evaluation of our experimental conditions, as the data are marginally affected by systematic errors such as dark counts, and we are considering samples much smaller than those collected to estimate $P_0$ and $P_1$ in the calibration step. The results are reported, for a vector of $N=100$ generated outcomes, in the histograms of Fig.\ref{fig:probs} for different choices of scenarios, considering both instances in which the probe is associated to a bosonic or a fermionic bath, in accordance to the fact that the error $\delta$ is symmetrised.

For each simulated time $\tau$ we indicate with different colours the fraction of events in which the bath has been correctly identified (blue) or mistaken (red) by following the Bayesian decision rule explained above - now with the actual values of $N_0$ and $N_1$, rather than their expected ones. The observed behaviours qualitatively mirror the errors in Figs.\ref{fig:error} and \ref{fig:horror}. Since $N_0$ and $N_1$ are now random variables, the discrimination capability exhibits deviations from the expected case.

\section{Conclusions}\label{CONC} 
Statistical tagging \cite{PhysRevA.100.042327} and, more generally, bath discrimination, is a simple yet insightful instance of the possibility of indirectly probing an environment~\cite{PhysRevLett.122.030403,
Brunelli2011qubit, PhysRevLett.114.220405, PhysRevA.96.012316, Campbell_2018, Kiilerich2018dynamical, PhysRevA.77.022111,Pekola,Aspelmeyer,Prokof_ev_2000,DePasquale,PhysRevA.97.012126}.
In this setting, information about the bath structure are retrieved via measurements on a quantum probe which has interacted with the bath up to a selected measurement time $\bar{t}~.$ %
This approach reveals how different properties of the bath affect the nature of the optimal discrimination procedures.
%
This is clear in the tagging context presented here:
a thermal bath has an unknown statistics - fermionic or bosonic - that we want to guess, with the additional information of knowing the respective temperatures - $1/\beta_f$ and $1/\beta_b$ - associated to the two bath instances. 
Here the quantum nature of the problem is manifested both in the statistical properties of the bath and in the coherence of the single-qubit probe. 
For input energy eigenstates, our inspection has revealed a transition between temperature regimes in which either equilibrium - $ \bar{t} \rightarrow \infty $ - or non-equilibrium states - $ \bar{t} < \infty $ - are optimal. 
Such behavior has been illustrated both theoretically and in a linear-optical simulation. 
States with quantum coherence, instead, do not display such transition - i.e. non-equilibrium conditions are generally optimal - and their inclusion allows to reach the best discrimination capability.\\
Extensions of this work may concern baths with richer features, such as very large baths presenting squeezing or, to the other extreme, small environments, entailing more involved treatments.
There, we can expect coherence properties of the probe to become even more relevant, thus adding richness and complexity to the observable phenomenology. \\

D.F. and V.G. acknowledge support from PRIN 2017 ``Taming complexity with quantum strategies".
During the completion of this manuscript, the authors became aware of a related work in preparation by L. Mancino et al.
``Non-equilibrium readiness and accuracy of Gaussian Quantum Thermometers" also dealing with metrological
tasks by means of indirectly measuring environments via quantum probes.

%

\begin{appendices}

\begin{widetext}

\subsection{Excited input state} \label{appendix-excited}
 Equation~(\ref{eq:hel}) clearly shows that the minimal values of HEP are achieved when 
 $\| \rho_b(t)-\rho_f(t)\|_1$ gets maximum.
From Eq.~(\ref{MAXI})  it follows that for the case of excited input state, i.e. for $a_z(0)=1$, 
this quantity can be expressed as
\bg || \rho_b(t) - \rho_f(t)||_1 = D(t,x,y) := |  (1- e^{- \gamma t})(1+x) - (1- e^{- \frac{\gamma t}{y}})(1+y)  |, \label{app1} \eg
where, for ease of notation, we 
introduced
\begin{eqnarray} x:=- a_z^{(f)}(\infty) = \tanh(\beta_f \omega/2)\;,   \label{xy} 
\qquad y:= -a_z^{(b)}(\infty) =\tanh(\beta_b \omega/2) \;, \end{eqnarray} 
and explicited
$\gamma_f=\gamma$, $\gamma_b=\gamma/y$.
Studying the Eq. (\ref{app1}) as a function of $ t$ we can infer which instant is optimal to perform a single measurement
for discriminating between the two hypotheses.

As first, we notice that $D(t,x,y)$ nullifies at $t=0$ (obviously) and at most in another point, since by solving
$D(t,x,y)=0$ we have
\bg  \frac{1+x}{1+y} = \frac{1- e^{- \frac{\gamma t}{y}}}{1- e^{- \gamma t}}. \label{app2} \eg
The unicity of the solution can be argued using the monotonicity of the $r.h.s.$ of Eq. (\ref{app2}).
Notice that the other solution (at $ t =0$) cannot be obtained from Eq. (\ref{app2}) since we divided by $(1-e^{-\gamma  t})$
that nullifies in that case.
The first derivative of $D( t,x,y)$ with respect to time reads
\bg  D'(t) = \gamma {\rm sign}[(1-e^{-\gamma  t})(1+x) -  (1- e^{-\frac{\gamma t}{y}})(1+y)] ((1+x)e^{-\gamma t} - (1+y) y^{-1} e^{-\frac{\gamma t}{y}}), \eg
that clearly nullifies in the long time limit $\gamma t \rightarrow \infty$.
To find other zeroes of $D'$ we have to solve the following equation
\bg \frac{1+x}{1+y^{-1}} = e^{-\gamma t(y^{-1} -1)}, \label{app3} \eg
that can have at most one solution since the $r.h.s.$ is a strictly decreasing function.
In addition, it is possible to prove that, calling $ t_1$ and $ t_2$ the zeores at finite time respectively of $D( t,x,y)$ and 
of its first derivative, we have $t_1 \geq t_2$.
Indeed they satisfy the two equations (\ref{app2}) and (\ref{app3}) from which we derive
\bg  \frac{e^{-\gamma t_2 (y^{-1}-1)}}{y}=  \frac{1- e^{- \frac{\gamma t_1}{y}}}{1- e^{- \gamma t_1}}. \label{app4}  \eg
Now we can use the following inequality:
$\frac{e^{-\gamma t_2 (y^{-1}-1)}}{y} \leq 
 \frac{1- e^{- \frac{\gamma t_2}{y}}}{1- e^{- \gamma t_2}}$,
from which $ t_1 \geq  t_2$ can be argued using the decreasing properties of both sides of Eq. (\ref{app4}).

As a last step we want to study the behaviour of the zeroes in the parameters $x,y$.
It is straightforward to verify that Eq.~(\ref{app1}) has no solutions if $x <y$, since there is 
no crossing between the bosonic and fermionic evolutions in this case.
Notice that, following the definition of $x$ and $y$, this last condition is equivalent to require the inverse temperature $\beta_f$ in the fermionic case to
be lower than the one in the bosonic case $\beta_b$.
Notice instead that the Eq.~(\ref{app3}) always nullifies once, independently from the value of $x,y$.
In conclusion, we have two possible qualitative trends for the trace norms (\ref{app1}):

\begin{enumerate}
 \item If $\beta_f < \beta_b$, $D( t)$ starts from $0$ and never nullfies again. The derivative
 of $D$ is zero once, in such a way that there is one single maximum. This case includes the 
 analysis done in \cite{PhysRevA.100.042327} in which $\beta_f = \beta_b$ was considered.
 \item If $\beta_f > \beta_b$, $D(t)$ reachs a maximum in $t_2$, then decreases to 
a point $t_1$ in which attains the value $0$. 
After $t_1$ $D(t)$ starts increasing again and remain monotonous when going to infinity.

\end{enumerate}

\subsubsection{Analysis of the critical point}
\label{appendix-criticality}

For the sake of characterizing the optimal measurements, we should find the maxima of the trace norm
studied in the previous section.
In the case $\beta_f < \beta_b$ there is only one maximum, and the measurement should be clearly done in the instant of time 
associated to that maximum.
In the case $\beta_f > \beta_b$ the intermediate maximum could be both greater or lesser than the value attained by $D(t)$
at infinitely long times.
In the following we will show that, again in dependence on the values assumed by the inverse temperatures 
$\beta_f$ and $\beta_b$, either one or the other strategy can be the best one.

To give a clear formulation to this question from a formal point of view, let us define the 
following function 
\bg  g(t,x,y) = D(t) - \lim_{t \rightarrow \infty} D(t). \eg
The zeroes of the function defined above correspond to the points in which $D(t)$ attains the same
value as it does at $t= \infty$.
Thus, if the equation $g(t,x,y)=0$ has no solutions, the absolute maximum is clearly located at $t \rightarrow 
\infty$.
Otherwise, given the properties of $D(t)$ enumerated in the previous section, the function $g(t,x,y)$ can 
have at most two zeroes, depending on the values of $x$ and $y$.
In this last case, the maximum is not located at $t = \infty$, since this last point is equal to at most two 
other values that the function $D(t)$ attains at finite time. 
It is also understood that the points
in which $g(t,x,y)$ has a unique zero (that will be referred from now as {\it critical}) are the ones in which 
$D(t)$ has two absolute maxima (identical in value).
If we fix the value of $y$ to some value $\bar{y}$, we have that in a critical point the solution $x_c(t, \bar{y})$ of
$g(t,x,\bar{y})=0$ must be such that $\frac{\partial}{\partial t} x(t, \bar{y}) =0$ evaluated in the critical point (this last property is 
derived from the regularity of $g$ and the definition of critical point, in which the Eq. $g(t,x,y)=0$ passes 
from having zero to two solutions).
We can then derive, using the definition of $g$, the following set of equations for the critical point

\bg  (2- e^{-\gamma t})(1+x) - (2- e^{- \frac{\gamma t}{\bar{y}}})(1+\bar{y}) =0, \\
        e^{-\gamma t} (1+x) - e^{- \frac{\gamma t}{\bar{y}}} \bar{y}^{-1} (1+\bar{y})= 0,
        \label{critical-curve}\eg
that replacing $\gamma t$ with $\tau$ gives the Eq. (\ref{critical-curve2}) of the main text.
For instance, choosing $\bar{y}= 1/2$ we have $x_c=  \frac{2 \sqrt{2} -1}{\sqrt{2}+1}$ and 
$t_c= \log(\frac{\sqrt{2} +1}{\sqrt{2}}).$        
Then, if we choose $x<x_c$ the better strategy is to measure at finite time, while if $x>x_c$ the measurement at the steady
state is the optimal one.
%


{\color{blue}  

}

\subsection{Discrimination with generic pure input states}
\label{app:generic}
Here we proceed with an analytical analysis of the HEP functional $H(t;a_z(0))$ defined in  Eq.~(\ref{QUESTA}).

  \subsubsection{Loss of susceptibility under non-coherent inputs} \label{app:generic1}
The worst  discrimination scenario is attained when HEP reaches its maximum value $1/2$: when this happens
the probability of error is maxima and we cannot recover information on the nature of the bath from the state of $A$.
From Eq.~(\ref{eq:hel}) this happens when $\| \rho_b(t)-\rho_f(t)\|_1=0$, i.e. when the two trajectories intercept. 
From Eq.~(\ref{QUESTA}) we observe that this can only occur when
\begin{eqnarray} 
\left\{ \begin{array}{l} 
 \label{ffg1} 
(e^{-\gamma_f t/2}- e^{-\gamma_b t/2})^2 (1- {a}^2_z(0))  = 0\;, \\ \\
(e^{-\gamma_f t}- e^{-\gamma_b t}) {a}_z(0) + a_z^{(f)}(\infty) (1- e^{-\gamma_f t})  -  a_z^{(b)}(\infty)  (1- e^{-\gamma_b t})  =0  \;. 
\end{array} \right. 
\end{eqnarray} 
However setting $|{a}_z(0)|<1$, i.e. allowing the input state of $A$ to be a non trivial superposition of the energy
eigenstates, this corresponds  to 
\begin{eqnarray} 
\left\{ \begin{array}{l} e^{-\gamma_f t/2} = e^{-\gamma_b t/2}\;, \\  \\
 (a_z^{(f)}(\infty)  -  a_z^{(b)}(\infty))  (1- e^{-\gamma_b t}) =0 \;, 
 \end{array} \right. 
\end{eqnarray} 
which can only be fulfilled for $t=\infty$ and  $\beta_f= \beta_b$. 
On the contrary setting ${a}_z(0)= \pm1$ (i.e forcing the probe to be in one of the two eigenstates of the system), 
the system~(\ref{ffg1}) reduces to a single equation 
\begin{gather} 
\pm (e^{-\gamma_f t}- e^{-\gamma_b t}) + a_z^{(f)}(\infty) (1- e^{-\gamma_f t})  -  a_z^{(b)}(\infty)  (1- e^{-\gamma_b t})  =0  \;,
\end{gather} 
which, depending on the specific values of $\beta_b$ and $\beta_f$ may allow for non trivial $t>0$ solutions, i.e. 
indicating a loss of susceptibility of the probe.

\subsubsection{Full optimization} \label{app:generic2}

We are interested in determining the minimum value of Eq.~(\ref{QUESTA}) 
 with respect to all possible inputs (i.e. all possible choices of $a_z(0) \in [-1,1]$) and
all possible times $t\geq 0$. According to (\ref{eq:hel}) this is formally equivalent to determining the maximum of
$\| \rho_b(t)-\rho_f(t)\|_1$ which in this case is given by the function 
\begin{eqnarray} 
\| \rho_b(t)-\rho_f(t)\|_1 = D(t;a_z(0))
\label{eq:d2-pure}
 &:=&    \Big\{ [(e^{-\gamma_f t}- e^{-\gamma_b t}) {a}_z(0) +  a_z^{(f)}(\infty) (1- e^{-\gamma_f t})  -  a_z^{(b)}(\infty)  (1- e^{-\gamma_b t})  ]^2 \nonumber \\
&&+
(e^{-\gamma_f t/2}- e^{-\gamma_b t/2})^2 (1 - {a}^2_z(0)  ) \Big\}^{1/2} \;.
\end{eqnarray} 
The best way to approach the problem seems to first optimize with respect to $a_z(0)$ and then maximize with respect to $t$.
We call $(\bar{t}, \bar{a}_z(0))$  the point where the maximum value of $D^2(t;a_z(0))
$ is attained.

Let us fix $t$ and rewrite $D^2(t;a_z(0))
$  as a parabola in ${a}_z(0):$
\begin{eqnarray}
\label{parabola}
D^2(t; a_z(0))&=&f_-^2(f_+^2-1) a_z^2(0) + 2 A f_- f_+ a_z(0)+f_-^2 + A^2~,
\end{eqnarray} 
with 
\begin{eqnarray} 
f_{\pm}:= e^{-\frac{\gamma t}{2}}\pm  e^{-\frac{\gamma t}{2y}}\;, \qquad 
A:=-x (1-e^{-\gamma t})+y(1-e^{-\frac{\gamma t}{y}})~,
\end{eqnarray}
where we used (\ref{xy}) to express the dependence upon $\beta_b$ and $\beta_f$. 
Since $a_z(0)\in [-1,1]~,$
$\bar{a}_z(0)$ is either one of the extrema -1 and 1 or the abscissa of the vertex $V=A f_+/[f_-(1-f_+^2)]$ of the parabola (\ref{parabola}).
The condition for the vertex to be the maximum is that the parabola is concave down and that the abscissa of the vertex falls strictly inside the interval $]-1,1[~:$
\begin{eqnarray}
\label{vertex-condition}
\frac{f_-}{f_+}(1-f_+^2) > \vert A \vert \Leftrightarrow \bar{a}_z(0)=V \in ]-1,1[~.
\end{eqnarray}

On the other hand, its violation imposes that the maximum is one of the extrema depending on the sign of A
\begin{eqnarray}
\label{no-vertex-condition}
\frac{f_-}{f_+}(1-f_+^2) \leq \vert A \vert  \Leftrightarrow \bar{a}_z(0)={\rm sign}[A]~.
\end{eqnarray} 
The equation above holds for $A\neq 0$, when $A=0$ the points $a_z(0)=1$ and $a_z(0)=-1$ are two equivalent maxima (still provided that the function 
is concave up).

Eventually, we have to find the maximum among 
$D^2(t_1; V_{t_1})$,  
$D^2(t_2; {1})$,
$D^2(t_2;{-1})$
for all $t_1$ satisfying inequality~(\ref{vertex-condition})
and $t_2$ satisfying inequality~(\ref{no-vertex-condition}). 
The explicit values of the three quantities above can be computed from Eq.~(\ref{parabola}) and read:
\begin{equation} \label{vitt1} 
D^2(t_1;V_{t_1}) = \frac{f^2_-(t_1) (f^2_+(t_1)-1)  - A^2(t_1)}{f^2_+(t_1)-1}\;;  \end{equation} 
\begin{equation} \label{vitt2} 
D^2(t_2;{1}) =  (f_+(t_2)f_-(t_2) + A(t_2))^2\;;  \end{equation} %
\begin{equation} \label{vitt3} 
D^2(t_2;{-1}) = (f_+(t_2)f_-(t_2) - A(t_2))^2 \;.
\end{equation}
Such maximization procedure yields the point $(\bar{t}, \bar{a}_z(0))$ we were searching for fixed $x$ and $y$.
However, notice that in general (for both the cases in which the concavity is 
up and down) the sign of A determines the sign of $\bar{a}_z(0)$
\begin{eqnarray}
{\rm sign}[\bar{a}_z(0)]={\rm sign}[A]~,
\end{eqnarray}
implying that in the region 
\begin{eqnarray}
y\geq x \Rightarrow
\bar{a}_z(0)>0~.
\end{eqnarray}
Moreover, condition (\ref{vertex-condition}) cannot be satisfied for $t$ sufficiently close to 0 such that
\begin{eqnarray}
\label{short-time}
\frac{\gamma t/2}{ \log\left(\frac{1}{1-\exp(-\gamma t/2)}\right)}\leq y
\Rightarrow \bar{a}_z(0)={\rm sign}[A]~.
\end{eqnarray}

\end{widetext}

\end{appendices}

\end{document}